\documentclass[a4paper]{jpconf}
\usepackage{graphicx}
\usepackage{float}
\usepackage{amsmath}
\usepackage{amsthm}
\usepackage{amsfonts}
\usepackage{citesort}

\begin{document}

\title{Dynamic analysis and control PID path of a model type gantry crane}

\author{P. A. Ospina-Henao$^{1,2}$, Framsol L\'opez-Suspes$^{1,2}$}
\address{1. Grupo de Investigaci\'on en Ciencias B\'asicas y Aplicadas. Departamento de Ciencias B\'asicas, Universidad 
Santo Tom\'as, Carrera 18 No. 9 - 27 PBX 6800801, Bucaramanga, Colombia.\\
2. Grupo de Investigaci\'on en Relatividad y Gravitaci\'on,
Escuela de F\'isica, Universidad Industrial de Santander, A. A. 678,
Bucaramanga, Colombia.}
\ead{paospina@uis.edu.co, framsol.lopez@ustabuca.edu.co}

\begin{abstract}
This paper presents an alternate form for the dynamic modelling of a mechanical system that simulates 
in real life a gantry crane type, using Euler's classical mechanics and Lagrange formalism, which allows find the equations
of motion that our model describe. Moreover, it has a basic model design system using the SolidWorks software, based on the
material and dimensions of the model provides some physical variables necessary for modelling. In order to verify the theoretical
results obtained, a contrast was made between solutions obtained by simulation in SimMechanics-Matlab and Euler-Lagrange equations
system, has been solved through  Matlab libraries for solving equation's systems of the type and order obtained. The force is
determined, but not as exerted by the spring, as this will be the control variable. The objective is to bring the mass of the 
pendulum from one point to another with a specified distance without the oscillation from it, so that, the answer is overdamped. 
This article includes an analysis of PID control in which the equations of motion of Euler-Lagrange are rewritten in the state 
space, once there, they were implemented in Simulink to get the natural response of the system to a step input in $F$ and then 
draw the desired trajectories. 
\end{abstract}

\section{Introduction}
There are many practical problems that require control system to be solved properly. Many of these problems involve highly 
complex dynamic systems, which often include nonlinearities, but can be approximated by more simple representations, i.e. the gantry type container cranes, used for the loading and unloading of ships in ports. The mass of the container, suspended 
from a trolley wire that carries along a rail, it presents a similar dynamic to the problem of the carriage and pendulum. 
This problem was configured as a case of an underactuated system: is necessary to control two variables, the carriage position 
and the mass by a single input signal, the force applied to the carriage. In addition, there are strict performance requirements: 
high response speed (cannot waste time in loading and unloading operations) and absence of oscillations in the position of the mass 
(the back and forth motion of the container can be dangerous). Container cranes are formidable machines typically employed
in port transhipment hubs for cargo movement whose efficiency, directly linked to the crane effective speed, strongly influences 
the productivity of the overall terminal. To optimize the crane performance in terms of minimization of the working operation time,
the accurate characterization of the dynamic response of containers during the lifting/lowering maneuvering phase is an essential
requirement. Moreover, harbors are often located in windy areas, characterized by the occurrence of gust fronts that can affect the
cranes working efficiency and, sometimes, compromise the safety of workers.

In the literature, several studies on container cranes dynamics are available and different modeling approaches (\textit{e.g.}, 
planar or three-dimensional models with elastic/rigid hoisting cables) are proposed to investigate their behavior. 
A comprehensive classification of these models is given in \cite{Abdel} where applications and limitations of the different 
theories are discussed. In order to investigate the dynamic behavior of container cranes, in recent years several models have 
been developed and control strategies have been proposed to overcome the potential issues arising from uncontrolled large-amplitude
oscillations. Studies of container crane pendulations were first conducted in \cite{Jones} where the
dynamic behavior of simply suspended objects was investigated and an extensive review of the theory was presented. In the same work,
a control strategy, based on programmed acceleration/deceleration phases of the trolley motion, was proposed and numerical results
were compared with experiments performed on a scale model. Payload oscillations may also be excited, as in ship-mounted
cranes, from the sea wave motion that can induce oscillations of the ship and, consequently, of the suspended container. 
Wave induced oscillations were described in \cite{Henry} where the planar motion of a one degree of freedom container crane model
was studied and a delayed position feedback controller was proposed.

Numerical simulations, aimed to study the effects of non linearities arising from mechanical friction and air resistance, were
performed in \cite{Liu} where the results were also experimentally validated. In \cite{Cartmell}-\cite{Morrish}, modeling of a 
variety of pendular structures (such as gantry cranes) was proposed and suitable control strategies were thoroughly examined by 
highlighting the computational implications of the models on the control implementations. A 1/10 laboratory scale model of a
gantry crane was employed in \cite{Cartmell} to test the performance of a feedback linearized control method, whereas 
in \cite{Morrish} single- and multi-cable models of gantry cranes were discussed.

The main purpose of controlling an underactuated crane system is transporting the payload in a precise location.
However, it is very difficult due to the fact that the payload can exhibit a pendulum-like swinging motion. Various
attempts in controlling cranes system based on open loop and closed-loop control system have been proposed. For example, open 
loop time optimal strategies were applied to the crane by many researchers \cite{Manson}-\cite{Auernig}. For example, PD 
controllers has been proposed for both position and anti-swing controls \cite{Omar}-\cite{Lee}. However, the performance of the
controller is not very effective in eliminating the steady state error.

The dynamics of the car-pendulum type gantry crane is more complex than the single pendulum-type overhead crane, so the system 
analysis and control algorithm design also is more difficult. In this paper, the gantry crane that exhibits car-pendulum dynamics
is investigated by using the Euler-Lagrange equations \cite{Meirovitch}-\cite{Goldstein}. 
The system model is built and the system properties are analyzed. An adaptive
PID control algorithm is proposed for the car-pendulum type gantry crane and the simulation
results demonstrate the system dynamics and the effectiveness of the proposed control algorithm. The  PID controller is the most 
common  form of feedback.  It  was an  essential element of early governors and it became the standard tool when process control 
emerged in the 1940s. In process control today, more than 95\% of the control loops are of PID type, most loops are actually PI 
control.  PID controllers  are  today  found  in  all areas where control is used. The controllers come in many different forms. 
There are stand alone systems  in  boxes for one or a few loops, which are manufactured by the hundred thousands yearly.  
PID control is an important ingredient of a distributed  control  system. The  controllers  are  also  embedded  in  many
special purpose control systems \cite{Sawodny}. PID control is often combined with logic, sequential  functions,  selectors,  
and  simple  function  blocks  to  build  the complicated automation systems used for energy production, transportation, and 
manufacturing.  Many  sophisticated  control  strategies,  such as model predictive control, are also organized hierarchically. 
PID control is used at the lowest  level; the multivariable  controller  gives  the setpoints to the controllers at the lower level.

The remainder of this paper is organized as follows. The dynamics of an actual configuration of a gantry crane transporting
distributed mass payloads, and we obtain the Euler-Lagrange equations for the model, is presented in Section 2. Next, in 
section 3, we presented the numerical model from based in the Solidworks and the comparative graphs in the analytical and 
the numerical solutions. In the Section 4, we presented the theory of small oscillations, we verify the natural frequencies and
normal coordinates is validate. In the Section 5, the PID control of the system is described above and experiments on a gantry
crane and the controller design procedure is presented. Block diagram of the system and the two variables which require control we 
presented. Finally, in Section 6, we present the conclusions of our main results.

\section{Symbolic model}
The crane type modeling system started from a model in which intervenes a spring at the point where the force is applied 
for the movement of the carriage on the rail, as shown in Figure 1 and the real gantry crane as shown in the Figure 2.
%\begin{figure}[H]
%\includegraphics[width=14pc]{figs/imagen_sistema.png}\hspace{2pc}%
%\begin{minipage}[b]{14pc}\caption{\label{label}Initial system image.}
%\end{minipage}
%\end{figure}

\begin{figure}[h]
\begin{minipage}{14pc}
\includegraphics[width=15pc]{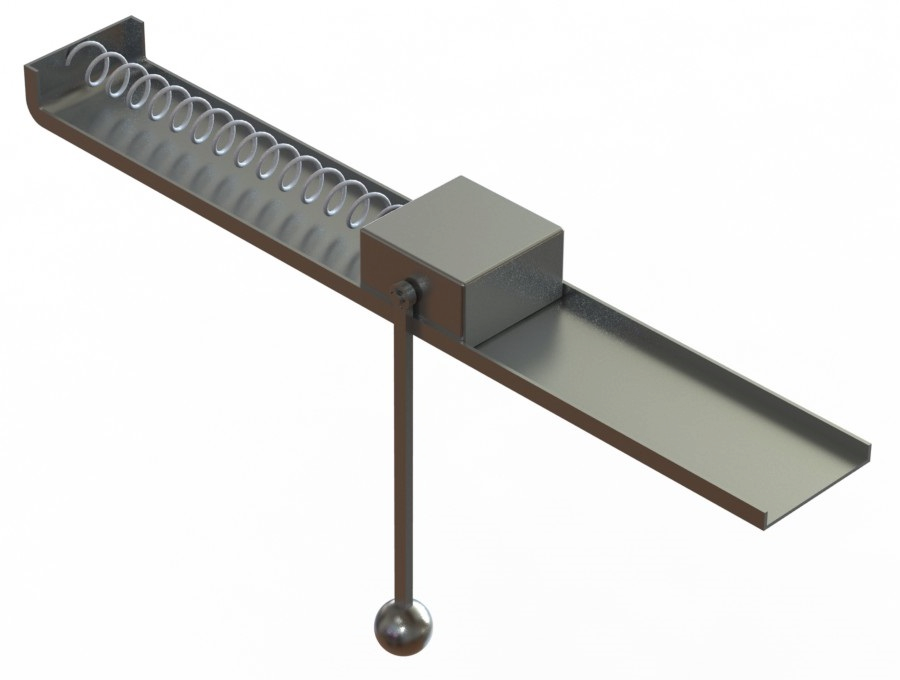}
\caption{\label{label}Initial system image}
\end{minipage}\hspace{2pc}%
\begin{minipage}{14pc}
\includegraphics[width=17pc]{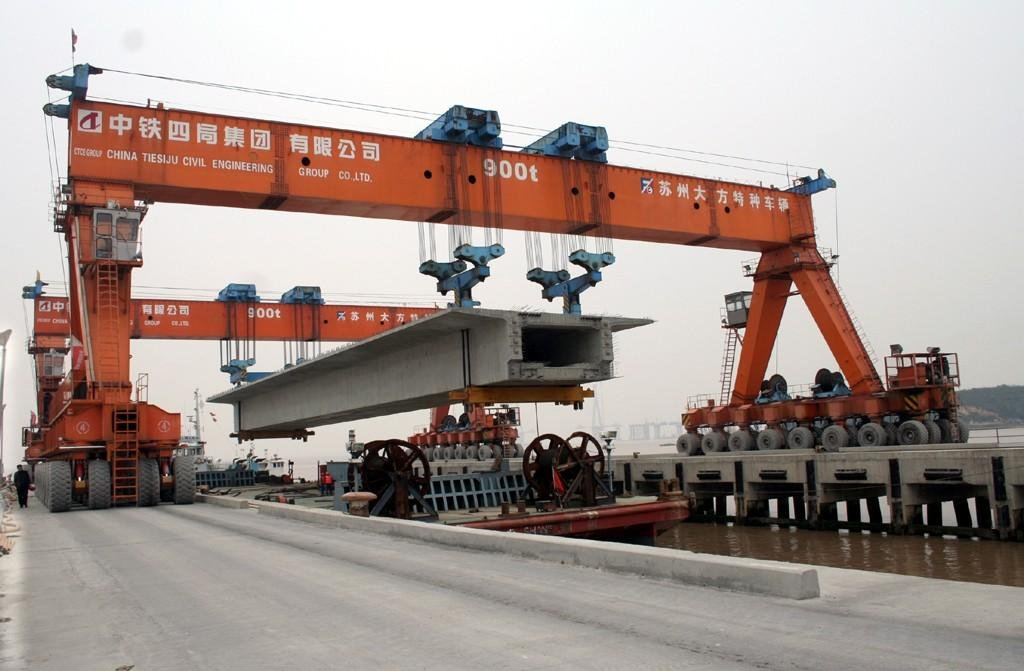}
\caption{\label{label}Gantry crane}
\end{minipage} 
\end{figure}

For this system the Lagrangian model was developed, in which initially the kinetic and potential energies are calculated.
For this model, was defined as $M$ is the mass of the carriage (object on the rail), $m_{1}$ the mass of the extreme of
the pendulum, $m_{2}$ the mass of the pendulum rod,  $\theta$ the angle between the vertical and the pendulum, the pendulum,  
$l$ the length and, $x$ the displacement of the carriage.

\subsection{{}Kinetic energy}
The kinetic energy is defined
\begin{equation}
T = \frac{1}{2}mv^2,
\end{equation}
where $m$ is the mass, $v$ velocity, and $T$ the kinetic energy of the body. In this case there are two bodies, the cart 
and the pendulum. The kinetic energy of the cart is,
\begin{equation}
T_1 = \frac{1}{2}M\dot{x}^2,
\end{equation}
the translational kinetic energy of the extreme of the pendulum,
\begin{equation}
T_2 = \frac{1}{2}m_1\left(l^2\dot{\theta}^2 + 2l\dot{\theta}\dot{x}\cos(\theta) + \dot{x}^2\right),
\end{equation}
the translational kinetic energy of the pendulum rod,
\begin{equation}
T_3 = \frac{1}{2}m_2\left(\frac{1}{4}l^2\dot{\theta}^2 + l\dot{\theta}\dot{x}\cos(\theta) + \dot{x}^2\right),
\end{equation}
the rotational energy of the pendulum rod is,
\begin{equation}
T_4 = \frac{1}{24}m_2\dot{\theta}^2l^2,
\end{equation}
and the rotational energy of the extreme of the pendulum is,
\begin{equation}
T_5 = \frac{1}{2}m_1\dot{\theta}^2\left(\frac{2}{5}r^2\right),
\end{equation}
where, $r$ is the radius of the sphere (extreme of the pendulum).

\subsection{{}Potential energy} 
This system stores potential energy in the spring and the pendulum, for spring,
\begin{equation}
U_1 = \frac{1}{2}Kx^2,
\end{equation}
where, $K$ is the elasticity constant of the spring. The potential energy of the pendulum is,
\begin{equation}
U_2 = -g\, l \cos(\theta)\left(m_1 + \frac{m_2}{2}\right),
\end{equation}
where, $g$ represents the value of the gravitational acceleration.

\subsection{{}Euler-Lagrange equations}
The Lagrangian of any system is defined,
\begin{equation}
\mathcal{L} = T - U,
\end{equation}
where, $T$ is the kinetic energy and $U$ the potential, therefore, for this model it is,
\begin{equation}
\begin{array}{llll}
	\mathcal{L} & = && T_1 + T_2 + T_3 + T_4 + T_5 - U_1 - U_2 \vspace{2mm}\\
	& = && \frac{1}{2}M\dot{x}^2 + \frac{1}{2}m_2\left(\frac{1}{4}l^2\dot{\theta}^2
	+ l\dot{\theta}\dot{x}\cos(\theta) + \dot{x}^2\right)\\
	&& + & \frac{1}{2}m_1\left(l^2\dot{\theta}^2 + 2l\dot{\theta}\dot{x}\cos(\theta) + \dot{x}^2\right)\vspace{2mm}\\
	&& + & \frac{1}{24}m_2l^2\dot{\theta}^2 + \frac{1}{2}m_1\dot{\theta}^2\left(\frac{2}{5}r^2\right) \vspace{2mm}\\
	&& + & g\, l \cos(\theta)\left(m_1 + \frac{m_2}{2}\right) - \frac{1}{2}Kx^2.
\end{array}
\end{equation}
The Euler-Lagrange equations are given by the following equation
\begin{equation}
\frac{d}{dt}\left(\frac{\partial \mathcal{L}}{\partial \dot{q}}\right) - \frac{\partial \mathcal{L}}{\partial q} = 0,
\label{eq:eqsL}
\end{equation}
where, $q$ represents the degrees of freedom (DOF) generalized, in this case, there are two DOF, $\theta$ and $x$, therefore, from 
equation (\ref{eq:eqsL}),
\begin{equation}
\frac{d}{dt}\left(\frac{\partial \mathcal{L}}{\partial \dot{\theta}}\right) - \frac{\partial \mathcal{L}}{\partial \theta} = 0
\label{eq:eqsL1}
\end{equation}

\begin{equation}
\frac{d}{dt}\left(\frac{\partial \mathcal{L}}{\partial \dot{x}}\right) - \frac{\partial \mathcal{L}}{\partial x} = 0.
\label{eq:eqsL2}
\end{equation}
From (\ref{eq:eqsL1}) and (\ref{eq:eqsL2}), the Euler-Lagrange equation, is obtained,
\begin{equation}
\ddot{\theta}l^2\,A + \left[\ddot{x}\,l\cos(\theta) + g\,l\sin(\theta)\right]\,B = 0
\label{eq:eqsLn1}
\end{equation}
\begin{equation}
\ddot{x}C - \left[\ddot{\theta}l\cos(\theta) - l\dot{\theta}^2\sin(\theta)\right] B + K\,x  =  0,
\label{eq:eqsLn2}
\end{equation}
where,
\begin{eqnarray}
A & = & \left(m_1 + \frac{m_2}{3}\right) + \frac{2\,m_1\,r^2}{5} \label{eq:A}\\
B & = & m_1 + \frac{m_2}{2} \label{eq:B}\\
C & = & m_1 + m_2 + M \label{eq:C}
\end{eqnarray}

\section{Numerical model}
To define the numerical model, the constant values are replaced,

\begin{center}
\begin{table}[H]
\centering
\caption{\label{jfonts}Constant values in the model.} 
\begin{tabular}{@{}l*{15}{l}}
\br
Constant&\;\;Value\\
\mr
$l$ & = & 0.2 [m] \\
$g$ & = & 9.81 $\mathrm{[m/s^2]}$ \\
$K$ & = & 100 [N/m] \\
$m_1$ & = & 0.088338025 [Kg] \\
$m_2$ & = & 0.022245336 [Kg] \\
$M$ & = & 0.548069759 [Kg] \\
$r$ & = & 0.02 [m]\\
\br
\end{tabular}
\end{table}
\end{center}
Substituting these values in equations (\ref{eq:eqsLn1}) and (\ref{eq:eqsLn2}),
\begin{equation}
0.003844\, \ddot{\theta} + 0.19514 \sin(\theta) + 0.01989\, \ddot{x} \cos(\theta)  =  0
\label{eq:numL1}
\end{equation}
\begin{equation}
\begin{array}{rcl}
0.6586\, \ddot{x} - 0.01989 \, \dot{\theta}^2 \sin(\theta) + 0.01989 \, \ddot{\theta} \cos(\theta) + 100\,x & = & 0
\end{array}
\label{eq:numL2}
\end{equation}
As can be seen, the equations (\ref{eq:numL1}) and (\ref{eq:numL2}) are coupled, that is to say, both depend on two variables 
to their second derivative, $\ddot{\theta}$, $\ddot{x}$, therefore, the initial conditions depend on the initial energy that 
shows this system, then, to find the initial conditions that can take, part of the potential energy is taken, which it is the
energy that accumulates the spring and the pendulum when $\theta$ is different from zero. The total potential energy of the
model is
\begin{equation}
\begin{array}{lll}
	U_T & = & U_1 + U_2 \vspace{2mm}\\
	& = & -g\, l \cos(\theta)\left(m_1 + \frac{m_2}{2}\right) + \frac{1}{2}K\, x^2,
\end{array}
\end{equation}
replacing the numerical values of the constants,
\begin{equation}
	U_T = -0.19514 \cos(\theta) + 50 \, x^2.
\end{equation}
There are infinite energy values that could take the system, however, only interested in those who do not turn the model into a
chaotic system, since it is assuming (until this moment) from conservative way, i.e., that no gain or loss of energy and therefore
the energy level that present, is due to the initial position of the pendulum and spring's compression. For spring it can be 
assumed values between $-0.1$ and $0.1$, and for the angular position of the pendulum $-\pi$ and $\pi$. From these values can be
graphed contours for energy values $0$, $0.1$ and $0.2$, considering that the reference system (zero potential energy) is located
in the center of rotation of the pendulum. In Figure 3, the potential energy levels for the system is shown, and in
Figure 4 the contours for three different energy levels.
\begin{figure}[H]
\begin{minipage}{14pc}
\includegraphics[width=17pc]{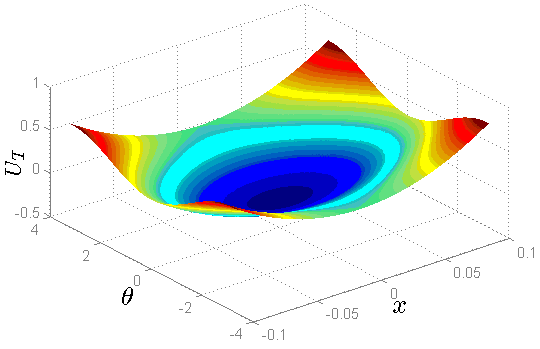}
\caption{\label{label}Energy level graph for the system}
\end{minipage}\hspace{6pc}%
\begin{minipage}{17pc}
\includegraphics[width=14pc]{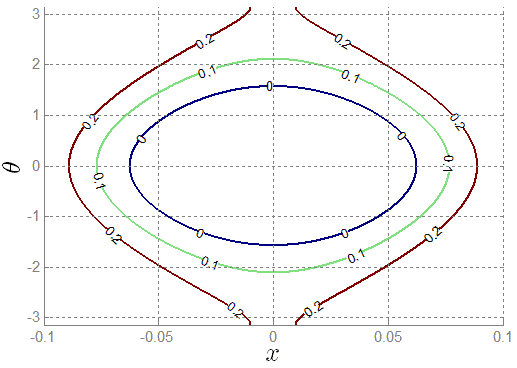}
\caption{\label{label}Contours for energy levels 0, 0.1 and 0.2}
\end{minipage} 
\end{figure}
Now, you can take the zero energy level, because it is stable curve is completely closed (as the graph in Figure 3), so that 
this curve the values of initial conditions are taken to solve the coupled differential equations (\ref{eq:numL1}) and 
(\ref{eq:numL2}). For convenience and ease, a value is taken as zero, $\theta = 0$, for which $x = 0.0624$. So that the initial
conditions are\\
\begin{center}
\begin{table}[H]
\centering
\caption{\label{jfonts}Initial conditions} 
\begin{tabular}{@{}l*{15}{l}}
\br
Conditions&\;\;Value\\
\mr
$x(0)$ & = & 0.0624\\
$\dot{x}(0)$ & = & 0\\
$\theta(0)$ & = & 0 \\
$\dot{\theta}(0)$ & = & 0 \\
\br
\end{tabular}
\end{table}
\end{center}

The numerical solution was performed in MATLAB, entering the equations (\ref{eq:numL1}) and (\ref{eq:numL2}), followed by the
next commands
\begin{verbatim}
V = odeToVectorField(eq1, eq2);
M = matlabFunction(V, 'vars', {'t', 'Y'});
sol = ode45(M, [0 10], [0.0624 0 0 0]);
\end{verbatim}
where \verb|eq1, eq2|, represent the numerical equations, and  \verb|sol|, the system solution, in this case MATLAB returns a 
structure (\verb|struct|), in which are contained a time vector and a matrix of four rows, each of these is a solution of the
variables, which order  is  $x, \dot{x}, \theta$ and $\dot{\theta}$ respectively. The solution for $\theta$ is shown in 
Figure 5 and $x$ for in Figure 6.\\

\begin{figure}[h]
\includegraphics[width=20pc]{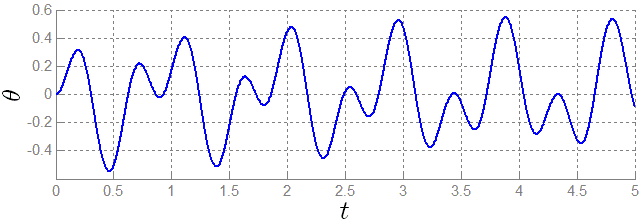}\hspace{2pc}%
\begin{minipage}[b]{14pc}\caption{\label{label}$\theta$ solution for the first 5 sec.}
\end{minipage}
\end{figure}

\begin{figure}[h]
\includegraphics[width=20pc]{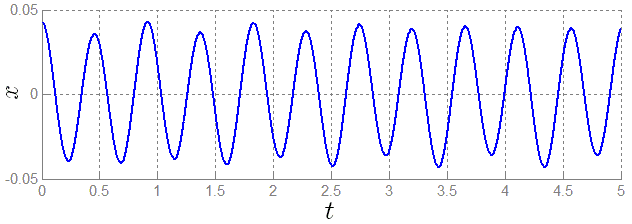}\hspace{2pc}%
\begin{minipage}[b]{14pc}\caption{\label{label}$x$ solution for the first 5 sec.}
\end{minipage}
\end{figure}

To validate the results obtained, the system was performed in SolidWorks and their movement was simulated considering 
the initial conditions, when comparing the results obtained analytically with the simulation of SolidWorks, a great  approach is observed, so 
you can have certainty about calculations performed from the Lagrangian's dynamics. Comparative graph is shown in Figure 7.

\begin{figure}[H]
\includegraphics[width=20pc]{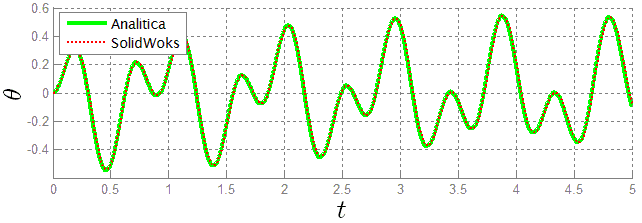}\hspace{2pc}%
\begin{minipage}[b]{14pc}\caption{\label{label}Comparative graph between analytical solution and simulation in SolidWorks.}
\end{minipage}
\end{figure}

\section{Oscillations}
The theory of small oscillations finds widespread physical applications in acoustics, molecular spectra, vibrations of mechanisms, and 
coupled electrical circuits. If the deviations of the system from stable equilibrium conditions are

small enough, the motion can generally be described as that of a system of coupled linear harmonic oscillators.\\
With the approximations for small angles
\begin{equation}
\sin(\theta) \approx \theta;\quad \cos(\theta) \approx 1; \quad \theta\,{\dot{\theta}}^2 \approx 0,
\end{equation}
the kinetic and potential energy are
\begin{equation}\label{EnergiaCinetica}
		T = \frac{1}{2}{\dot{\theta}}^2\left[m_1\,l^2 + \frac{7}{12}m_2\,l^2\right] + \frac{1}{2}{\dot{x}}^2\left[m_2 + M\right] + \dot{\theta}\,\dot{x}\,l\left[m_1 + \frac{1}{2}m_2\right]
\end{equation}

\begin{equation}\label{EnergiaPotencial}
		U = \frac{1}{2}K\,x^2 + \frac{1}{2}{\theta}^2g\,l\left(m_1+\frac{1}{2}m_2\right).
\end{equation}
It is obvious from their definition that the $U$ are symmetrical, that is, that $U_{ij} = U_{ji}$. A similar series expansion can be 
obtained for the kinetic energy. Since the generalized coordinates do not involve the time explicitly, the kinetic energy is a
homogeneous quadratic function of the velocities.
Expressing equations (\ref{EnergiaCinetica}) and (\ref{EnergiaPotencial}) matrix form
\begin{equation}\label{MatrizCinetica}
		\textbf{T} = 
		\begin{bmatrix}
				l^2\left(m_1+\frac{7}{12}m_2\right) & l\left(m_1+\frac{1}{2}m_2\right)\\
				\\
				l\left(m_1+\frac{1}{2}m_2\right) & m_2+M
		\end{bmatrix}
\end{equation}

\begin{equation}\label{MatrizPotencial}
		\textbf{U} = 
		\begin{bmatrix}
				g\,l\left(m_1+\frac{1}{2}m_2\right) & 0\\
				\\
				0 & K
		\end{bmatrix}
\end{equation}

where the matrix (\ref{MatrizCinetica}) and (\ref{MatrizPotencial}) meet the following two equations

\begin{equation}
		\textbf{T} = \frac{1}{2}\textbf{T}_{i\,j}\,\dot{\theta}\:\dot{x}
\end{equation}

\begin{equation}
		\textbf{U} = \frac{1}{2}\textbf{U}_{i\,j}\theta\,x.
\end{equation}

\subsection{Frequences of free vibration and normal coordinates}
To find the oscillation frequencies of the secular equation solves

\begin{equation}\label{EcuacionSecular}
		|\;\textbf{U} - {\omega}^2\textbf{T}\;| = 0,
\end{equation}
where, $\omega$ represents the eigenvalues, for this case, frequencies are solving the equation (\ref{EcuacionSecular}), with the 
obvious solutions
\begin{equation}\label{Omega1}
		{\omega}_1 = \sqrt{\frac{g\left(M+m_2\right)+K\,l+\sqrt{{g^2\left(M+m_2\right)}^2+K\,l\left(K\,l+4m_1g-2M\,g\right)}}{l\left(2M-2m_1+m_2\right)}}
\end{equation}

\begin{equation}\label{Omega2}
		{\omega}_2 = \sqrt{\frac{g\left(M+m_2\right)+K\,l-\sqrt{{g^2\left(M+m_2\right)}^2+K\,l\left(K\,l+4m_1g-2M\,g\right)}}{l\left(2M-2m_1+m_2\right)}}
\end{equation}
if the spring constant $K$ and mass $M$ were $0$, would not determine the $x$ variable, and the frequency ${\omega}_2$ would
\begin{equation}\label{Omega3}
		{\omega}_2 = \sqrt{\frac{g}{l}\frac{2m_2}{(m_2-2m_1)}}
\end{equation}
The equation (\ref{Omega3}) allows you to see clearly that corresponds to an oscillator
harmonic. Now if the masses $m_1$ and $m_2$ were equal to zero, the frequency $\omega_1$ was reduced to
\begin{equation}\label{Omega4}
		{\omega}_1 = \sqrt{\frac{K}{M}},
\end{equation}
which also corresponds to the frequency of a harmonic oscillator.
Through own and knowing the eigenvalues (natural frequencies) vectors $\xi$ normal modes of oscillation are calculated, according to this, we
\begin{equation}\label{Cond1}
		\left(\textbf{U}-{\omega}_2\textbf{T}\right)\textbf{a}_k = 0
\end{equation}

\begin{equation}\label{Cond2}
		{\textbf{a}^T}_k\textbf{T\,a}_k = 1,
\end{equation}
where, 
\begin{equation}
		\textbf{a}_k =
		\begin{bmatrix}
				a_{1\,k}\\
				a_{2\,k}
		\end{bmatrix}
		\quad with\;k = 1,2
\end{equation}

\begin{equation}
		\textbf{A} = \left[\textbf{a}_1\quad\textbf{a}_2\right].
\end{equation}
Here, $\textbf{A}$ is the matrix of eigenvalues and equation (\ref{Cond2}) is provided
orthogonality. The normal modes of oscillation are defined from the equation (\ref{Modos}), this is
\begin{equation}\label{Modos}
		\hbox{\boldmath $ \xi = A^T\,T\,\eta $ \unboldmath},
\end{equation}
with
\begin{equation}
		\hbox{\boldmath $ \eta $ \unboldmath} =
		\begin{bmatrix}
				\theta \\
				x
		\end{bmatrix}
\end{equation}
The normal modes of oscillation are
\begin{equation}
	\hbox{\boldmath $ \xi $ \unboldmath} =
	\begin{bmatrix}
		\frac{\sqrt{2}\:B}{2\:\sqrt{A}}\\\\
		-\frac{\sqrt{2}\:B}{2\:\sqrt{A}},
	\end{bmatrix}
\end{equation}
where $A$ and $B$ are defined as
\begin{equation}
	A = 2\,M\left[g^2+l^2{\omega_1}^4-2g\,l{\omega_1}^2\right]+2m_2g\left[g-l{\omega_1}^2\right]+2{\omega_1}^2m_1\,l\left[2\,g-l{\omega_1}^2\right]
\end{equation}

\begin{equation}
  B= 2Mx\left[g-l{\omega_2}^2\right]+m_2\left[2gx-lx{\omega_2}^2+\theta gl\right] + 2m_1l\left[x{\omega_2}^2+\theta g\right].
\end{equation}
We have spoken so far only of vibrations along the axis; 
there will also be normal modes of vibration perpendicular to the axis. The 
complete set of normal modes is naturally more difficult to determine than the
longitudinal modes, for the general motion in all directions corresponds to more
degrees of freedom.

\section{PID Control}

Many researchers use complicated nonlinear equations or even partial differential equations as a model. But, whatever model they use, the final 
ontrol law should be given as a function of the acceleration of the trolley or should be derived through the dynamics of trolley motion, 
in Hong \textit{et al.}\cite{Hong}. In \cite{Oliveira}, proposes a Gantry crane control simulation experiment exploring the concepts of 
half-cycle posicast feedforward control, and Proportional, Integrative and Derivative (PID) control. Two control objectives are considered: 
set-point tracking and load disturbance rejection. The set-point tracking objective in this problem consists in moving the suspended load from a 
initial position to a specified final position. In \cite{Singer}, present a new procedure for designing input shapers for gantry cranes was presented.
The procedure takes into account the unique properties of gantry cranes, such as, the single-mode dynamics, the known frequency range, and the 
standard deceleration period. The new shapers are of fixed duration, so that they give a constant deceleration period. Given the fixed shaper 
duration, the robustness of the shapers is maximized. The new shaping method was implemented on a gantry crane at the Savannah River Technology Center.
Experimental results show that the method greatly reduces residual oscillations and closely matches the theoretically predicted performance.\\
To control the system described above, the spring is deleted, which exerts a force on the mass placed on the rail. The force is determined, but not 
as exerted by the spring, since this will be the variable being monitored. The objective is to bring the mass of the pendulum from one point to another
with a specified distance it to oscillate, so the answer is overdamped. \\
In \cite{Ahmad}, this article then extended to incorporate a non collocated PID controller for control of sway angle of the pendulum. Implementation 
results of the response of the rotary crane system with the controllers are presented in time and frequency domains. The performances of the
control schemes are assessed in terms of level of sway reduction, rotational angle tracking capability and time response specifications.\\

To control the system described above, the spring is deleted, which exerts a force on the mass placed on the rail. The force is determined, but not as
exerted by the spring, since this will be the variable being monitored. The objective is to bring the mass of the pendulum from one point to another 
with a specified distance it to oscillate, so the answer is overdamped.

\begin{figure}[H]
\includegraphics[width=20pc]{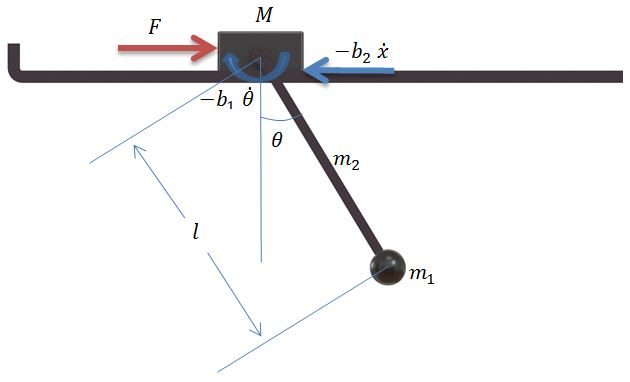}\hspace{2pc}%
\begin{minipage}[b]{14pc}\caption{\label{label}PID control model variables.}
\end{minipage}
\end{figure}
From the system described in Figure 8, the dynamic model is determined from the Euler-Lagrange equations for generalized 
force, this is
\begin{equation}
\dfrac{d}{dt}\left(\frac{\partial \mathcal{L}}{\partial \dot{q}_i}\right) - \frac{\partial \mathcal{L}}{\partial q_i} = Q_i, \qquad i = 1, 2
\label{eq:eqsLQ}
\end{equation}
where $q_1 = \theta$, $q_2 = x$, $Q_1 = Q_{\theta}$ and $Q_2 = Q_x$; these last two represent the generalized forces over the
system, which will no longer be conservative, due to the existence of an external force $F$ and friction is
assumed in the pendulum axis ($b_1 \dot{\theta}$) and carriage contact with the rail ($b_2 \dot{x}$), according to Figure 8. With
this variation, the new Euler-Lagrange equations, are
\begin{equation}
\ddot{\theta}\,D + \left[\ddot{x}\,l\cos(\theta) + g\,l\sin(\theta)\right]\,B = -b_1\dot{\theta}
\label{eq:EL1}
\end{equation}
\begin{equation}
\ddot{x}\,C + \left[\ddot{x}\,l\cos(\theta) + g\,l\sin(\theta)\right]\,B = F - b_2\,\dot{x}
\label{eq:EL2}
\end{equation}
where
\begin{equation}
D = \left(\dfrac{2}{5}m_1\,r^2\right),
\label{eq:D}
\end{equation}
and $B$, $C$ are the same terms in the equations (\ref{eq:B}) and (\ref{eq:C}), respectively.
To control the system, you must first find the state space,
\begin{equation}
\mathbf{Z} =
	\left\{
		\begin{array}{c}
			\theta \\
			\dot{\theta} \\
			x \\
			\dot{x}
		\end{array}
	\right\},
\end{equation}
such that the derivative of $\mathbf{Z}$ is
\begin{equation}
\dot{\mathbf{Z}} =
	\left\{
		\begin{array}{c}
			\mathrm{Z}_2 \vspace{2mm}\\
			\dfrac{l\,B\cos(\mathrm{Z}_1)\,\epsilon_1 + C\,\epsilon_2}{\epsilon_3} \vspace{2mm} \\
			\mathrm{Z}_4 \vspace{2mm} \\
			\dfrac{\left(l^2\,B + D\right)\,\epsilon_1 + \left(l\,B\cos(\mathrm{Z}_1)\right)\,\epsilon_2}{-\epsilon_3}
		\end{array}
	\right\},
	\label{eq:dZ}
\end{equation}
where
\begin{equation*}
\begin{array}{lll}
	\epsilon_1 & = & F - b_2\dot{x} + \dot{\theta}^2l\,B\sin(\mathrm{Z}_1), \vspace{2mm} \\
	\epsilon_2 & = & b_1\dot{\theta} + g\,l\,B\sin(\mathrm{Z}_1), \vspace{2mm} \\
	\epsilon_3 & = & l^2\,B^2\cos^2(\mathrm{Z}_1) - l^2\,B\,C - D\,C.
\end{array}
\end{equation*}
The equations in space of states (\ref{eq:dZ}) were implemented in Simulink to obtain the natural response of
the system to a step input in $F$ and subsequently draw the desired trajectory. In the scheme of Simulink,
the value for $b_1 = 0.1$, and $b_2 = 0.5$, was assumed; because the system is dimensionally small, the
input for the force $F$ was $0.1$, for a time of $2$ seconds. The natural response to these values are shown in Figure 9 and
Figure 10 for $\theta, x$ respectively.
\begin{figure}[H]
\begin{minipage}{14pc}
\includegraphics[width=20pc]{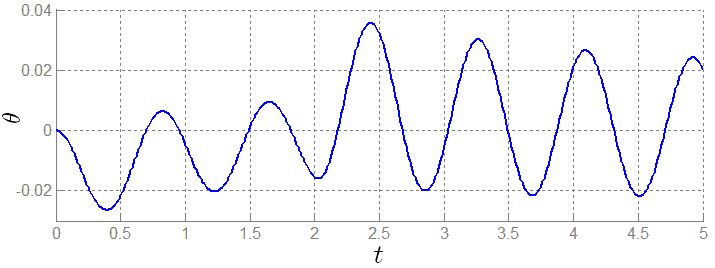}
\caption{\label{label}Output to an input $F = 0.1$, for $2$ seconds.}
\end{minipage}\hspace{6pc}%
\begin{minipage}{17pc}
\includegraphics[width=20pc]{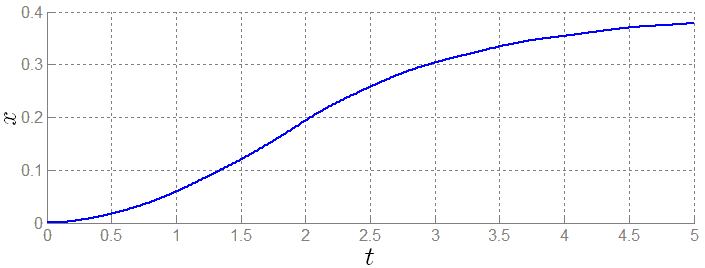}
\caption{\label{label}Output to an input $F = 0.1$, for $2$ seconds.}
\end{minipage} 
\end{figure}
A suitable trajectory for the variable $x$ is that in which the speed starts and ends at zero, the trigonometric
function $\tanh$, so that the function of the desired trajectory is,
\begin{equation}\label{TrayectoriaIdeal}
x^* = \frac{d}{2}\tanh\left[\alpha\left(t - \frac{T}{2}\right)\right] + \frac{d}{2},
\end{equation}
where $d$ is the distance to travel, $t$ represents time, $T$ the time and $\alpha$ a factor to vary the slope of the
curve. The function type proposed is more than twice differentiable, this ensures a continuous acceleration
along the way. Having defined the trajectory to $x$, these values are entered into equation (\ref{eq:EL1}), and
trajectories for $\theta$ y $\dot{\theta}$ are found. To the controller of the
variable $\theta$, the tuned values are
\begin{center}
\begin{table}[H]
\centering
\caption{\label{jfonts}Controller $\theta$} 
\begin{tabular}{@{}l*{15}{l}}
\br
Conditions&\;\;Value\\
\mr
Proportional & = & 3\\
Integral & = & 0.25\\
Derivative & = & 0.5 \\
\br
\end{tabular}
\end{table}
\end{center}
and for the variable $x$, we are
\begin{center}
\begin{table}[H]
\centering
\caption{\label{jfonts}Controller $x$} 
\begin{tabular}{@{}l*{15}{l}}
\br
Conditions&\;\;Value\\
\mr
Proportional & = & 0.01\\
Integral & = & 1.7\\
Derivative & = & 2 \\
\br
\end{tabular}
\end{table}
\end{center}

The Figure 11 shows the block diagram for this system, 
\begin{figure}[H]
\includegraphics[width=18pc]{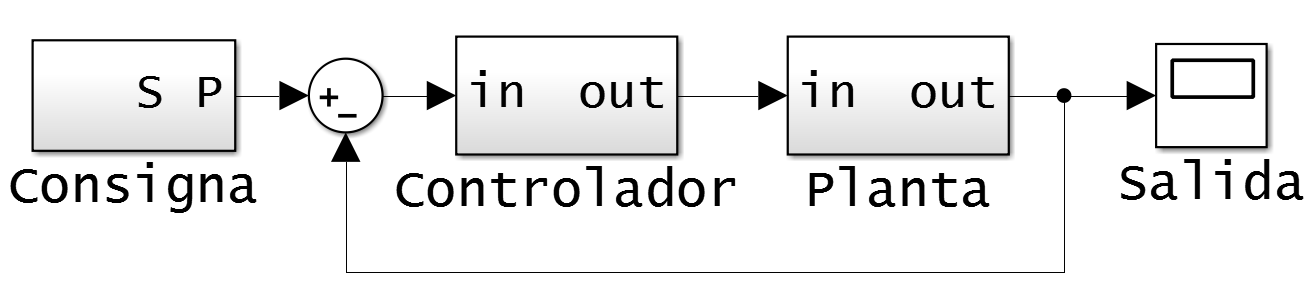}\hspace{2pc}%
\begin{minipage}[b]{14pc}\caption{\label{label}Block diagram of the system with the incorporated controller.}
\end{minipage}
\end{figure}
in which the controller is already implemented, which in turn has two PID, because they are two variables which require control. This diagram 
represents the general model of the system with the implemented controller, in which the $\it{Planta}$(Plant) block represents the
mechanical system, or the grantry crane. From this are taken readings of angular position of the pendulum and linear position of the car, 
as well as their respective speeds and accelerations. These output parameters are subtracted from the $\it{Consigna}$(Set point) and the error is 
determined (which is simply the subtraction of the desired output with the actual output), which becomes the PID controller input, sending a 
power signal to the engine that moves the system (Plant).

The output for $\theta, \dot{\theta}$ and $x, \dot{x}$ is registered in Figure 12 and Figure 13, respectively
\begin{figure}[H]
\begin{minipage}{14pc}
\includegraphics[width=20pc]{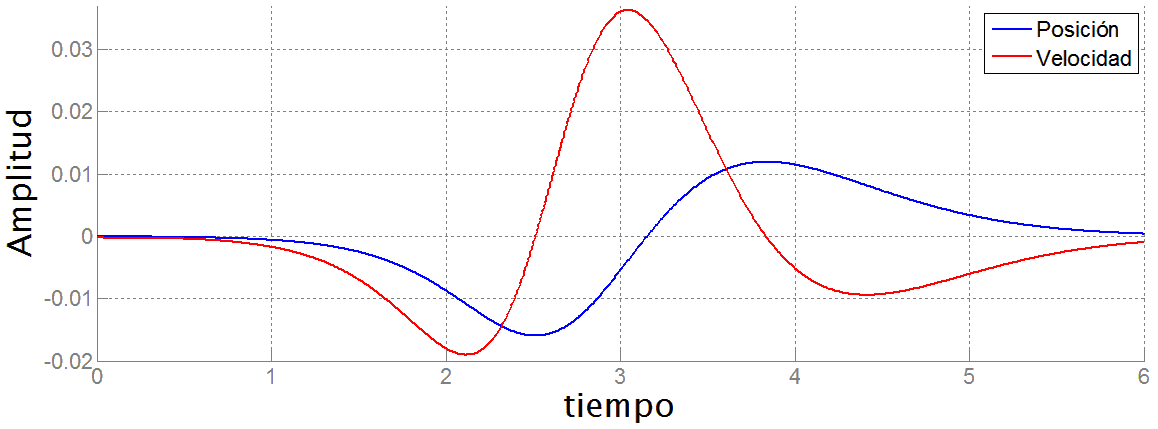}
\caption{\label{label}Output for $\theta$ and $\dot{\theta}$.}
\end{minipage}\hspace{6pc}%
\begin{minipage}{17pc}
\includegraphics[width=20pc]{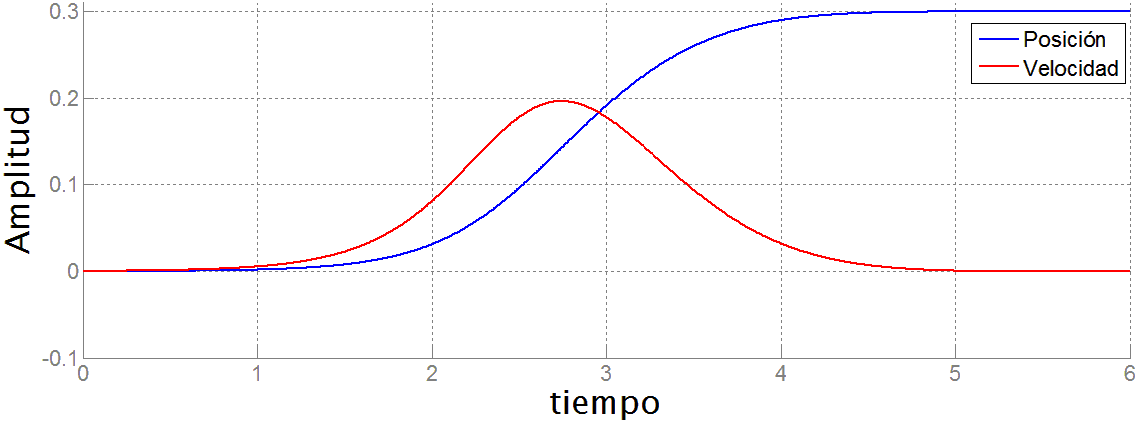}
\caption{\label{label}Output for $x$ and $\dot{x}$.}
\end{minipage} 
\end{figure}
As seen in the previous two graphs, variables were controlled according to the desired trajectory. In addition to the
equations included in Simulink, the CAD model was implemented in SimMechanics,
with the purpose of verify the controller system, which is made use of SimMechanics-link, which
exports SolidWorks CAD model to MATLAB. 
As for the position and velocity of the pendulum shown in Figure 12, it can be seen that the controller tries to keep the pendulum at 
its equilibrium point, It can be seen in the red line that the position has few oscillations, so that remains in cuasi-equilibrium.
The Figure 13 shows the position and speed of the cart under the
action of the controller, it is observed that the cart arrives at the
desired position by the trajectory defined in a non-abrupt way
and without going back. As for the speed, we see that this
increases when the cart is starting and decreases when it is
reaching its stationary zone, so that the car remains with a null
position error.

%\begin{figure}[H]
%\includegraphics[width=20pc]{figs/modelSpring.png}\hspace{2pc}%
%\begin{minipage}[b]{14pc}\caption{\label{label}System model in SimMechanics.}
%\end{minipage}
%\end{figure}

%\begin{figure}[H]
%\includegraphics[width=16pc]{figs/bloques2.png}\hspace{2pc}%
%\begin{minipage}[b]{14pc}\caption{\label{label}SimMechanics full diagram.}
%\end{minipage}
%\end{figure}
%The block diagram in SimMechanics is shown in Figure
%14, this figure shows the spring which initially was used to simulate the dynamic model that included,
%subsequently suspended for implementing the controller. The full diagram SimMechanics with the
%controller is shown in Figure 14. The block model of Fig. 15, contains the same block diagram of Fig. 14, omitting the spring, and
%blocks ${\it Consigna}$ and ${\it Controlador}$ are the same as implemented in Fig. 11, the variables were
%measured directly of the blocks ${\it Prismatic}$ and ${\it Revolute2}$.

\section{Conclusions}
The Euler-Lagrange formalism allows dynamic modeling of crane-type system in a simple way, thanks
to the formalism is based on energy principles. This paper has been presented step by step the dynamic
modeling of the proposed system, in addition to the respective simulation of the solution of the equations
and later validation of the results using a CAD model simulation designed in SolidWorks in the
extension of SimMechanics Matlab. This document differs from others through the use of
SimMechanics extension for simulation CAD model, this tool is simple to use and reliable, this is anadvantage for
simulations of 3D model since the creation of platforms is not necessary for their
simulation. The results were represented graphically and are successful, since two different ways came
to the same results with a visually low error between the two procedures.
Additionally, in this particular article analysis it is made as to the desired trajectories, such phase space
analysis is topologically correct, this being a good contribution to future dynamic studies, kinematic and
physical of the system. The development of this work requires a kind of dynamic modelling to study the
properties needed for monitoring tasks using techniques associated with control.  Consists in know thoroughly the system
under study, the context in which it is located, meet its internal dynamics, variables, its operation, its
components and the relationships between them, the maximum error to tolerate, alternate models are
known, the point(s) of desired operation, the operating range, etc. In addition, it must satisfy the
objectives and expectations of the dynamic to develop, among other things of the model.
The PID is a type of controller in state space that responds well to perturbations. In addition, this driver guarantees
us a null position error. The controller responds correctly, since it takes the carriage
to the desired position in a good way following the path defined in (\ref{TrayectoriaIdeal}).
Another step is the characterization of the variables and equations of motion. In this phase the dynamic
model of the proposed systems was studied, for which the Euler-Lagrange formalism was used because
it fits perfectly for the study of nonlinear dynamical systems. After having the motion equations of the
systems, it proceeded to solve the differential equations obtained, which are partial, of the second order,
and coupled. The phase control, PID control was used, with which we obtained a synthesis of the drivers
from the equations of motion but in the space of states, obtained in the characterization phase, it was
possible to represent dynamic systems as a whole of coupled differential equations of first order. In this
representation, known as representation in the state space, it shows the variation of a vector z, known as
state vector and contains the most important system information, as a function of the same z vector and
an input vector u. From this formulation, it was possible to synthesize laws nonlinear control.

\subsection{Acknowledgments}
Authors want to thank the financial support from Mathematical School from Universidad Industrial de Santander
(UIS) and Universidad Santo Tom\'as.

\section*{References}
%\bibliography{iopart-num}

%\end{document}

\end{document}